\documentclass[aps,preprint,showpacs]{revtex4}
\usepackage{graphicx}
\usepackage{amsmath}
\usepackage{epsfig}
\begin{document}

\title{Incoherent mid-infrared charge excitation and the high energy anomaly
in the photoemission spectra of cuprates}
\author{S. Cojocaru$^{1,2}$, R. Citro$^{1}$ and M. Marinaro$^{1}$}
\affiliation{$^{1}$ Dipartimento di Fisica ``E. R. Caianiello'' and C.N.I.S.M. , Universit%
{\`a} degli Studi di Salerno, Via S. Allende, I-84081 Baronissi (Sa), Italy}
\affiliation{$^{3}$ Institute of Applied Physics, Chi\c{s}in\u{a}u 2028, Moldova}

\begin{abstract}
On the basis of a semi-phenomenological model, it is argued that the high
energy anomaly observed in recent photoemission experiments on cuprates is
caused by interaction with an overdamped bosonic mode in the mid-infrared
region of the spectrum. Analysis of optical conductivity allows to connect
this excitation to the incoherent charge response reported for the majority
of high Tc materials and some other perovskites. We show that its large
damping is an essential feature responsible for the \textquotedblleft
waterfall\textquotedblright\ dispersion and linewidth of the spectral weight.
\end{abstract}

\maketitle

The \textquotedblleft high energy anomaly\textquotedblright\ (HEA) in the
angle resolved photoemission spectra of hole and electron doped cuprates has
been reported recently in a number of works \cite{Graf}-\cite{Pan}. The
anomaly is characterized by the presence of high energy kink in the
dispersion derived from angle resolved photoemission spectroscopy (ARPES) at
about $0.2-0.4$ eV below the Fermi surface followed by a puzzling
\textquotedblleft waterfall\textquotedblright\ structure at larger energies.
The latter is highly incoherent and follows an almost vertical dispersion
around the $\Gamma -$point extending till $0.9$ eV. When measured along the
main symmetry directions of the Brillouin zone such broad linewidth remains
roughly constant along the waterfall \cite{Graf, Pan}. The momentum
distribution of the spectral weight at a fixed energy above the kink
resembles a diamond shape (e.g. for $E=0.4$ in BISCO 2212 with the corner at
$K\simeq \left( \pi /3,0\right) $ and side center at $K\simeq \left( \pi
/4,\pi /4\right) $). At even higher energies the spectrum recovers the
parabolic dispersion reminiscent of the \textquotedblleft
bare\textquotedblright\ band. Several mechanisms responsible for the
observed behavior have been proposed, such as disintegration of the
quasiparticles into a spinon and holon \cite{Graf}, polaronic effects \cite%
{Xie}, spin fluctuations whose spectrum can reach the relevant energy scale
\cite{Valla, Macridin, Markiewicz}, coherence-incoherence crossover \cite%
{Pan, Chang, Byczuk}. However a controversy also exists in experimental
data, e.g. regarding the role of matrix element effects in ARPES (\cite%
{KordyukBorisenko, Alexandrov, Chang}). The fixed energy slices of the
incoherent tails reveal the existence of a \textquotedblleft
grid-like\textquotedblright\ structure when the spectral weight maxima are
aligned to the $\left( 0,1\right) ,\left( 1,0\right) $ directions and
suppressed in the diagonal directions. It was suggested that selfenergy
effects may therefore not be a cause of the anomaly which instead should be
related to the presence of one dimensional structure. On the other side some
experiments on magneto-optics spanning a broad frequency window (4000 cm$%
^{-1}$) are apparently inconsistent with magnetic excitation scenario \cite%
{LeeSegawa, DordevicHomes}.

In Ref. \cite{HwangNicol} the optical conductivity spectra have been
analysed by a method allowing to extract the electron selfenergy within
apparently general assumptions about the spectrum. However, no indication of
the HEA has been found, while the well known low energy kink (LEK) and the
above mentioned feature at $0.9$ eV are reproduced in good agreement with
ARPES. On the other side, the analysis based on fermion-boson concept in Ref.%
\cite{Norman} before the discovery of HEA has revealed that optical data do
indicate the existence of a high energy cutoff of the boson spectrum at
around $300$ meV. Also in several other earlier papers treating
electron-boson scenario it has been demonstrated that taking into account
the finite linewidth of the boson is important for the faithful description
of the ARPES spectra \cite{Cap, Carb, Dogan}. In particular, the LEK is well
reproduced if the linewidth of the boson (phonon) is about $5-10\%$ of its
energy. In this paper we show that the qualitatively new features of the HEA
can be accounted for by assuming the \textquotedblleft
overdamped\textquotedblright\ regime for the \textquotedblleft
boson\textquotedblright , i.e., that the linewidth of this excitation is
significantly larger than the characteristic energy. This assumption is
motivated by the ubiquity of such incoherent excitations in the mid-infrared
(MIR) frequency range of the optical spectra in cuprates as revealed, e.g.,
by Drude-Lorentz analysis. As explained below, although the bare frequency
of such excitation is larger than the high energy kink, the scale that
actually characterizes its effect on electron excitations falls in the
relevant energy range. We consider the selfenergy effects within a simple
model of electrons interacting with a bosonic mode in the form of damped
oscillator. The parameters of the oscillator are chosen to match the known
experimental data, in particular those from ARPES. As for instance, its
coupling as extracted from the slope of single particle spectrum in the
region preceding the HEA (e.g. for BISCO 2212 and LBCO is $\lambda _{c}$ $%
=1-1.5$), momentum and energy location of the observed structures etc. \cite%
{Valla, KulicDolgov}. Calculations show that the constraints set by the
model and experimental data are rather restrictive so that the choice of
parameters can not be reduced to simple fitting. For instance, it is not
possible to obtain a vertical dispersion by choosing a small damping
parameter and varying the others. We show that, indeed, the key component of
the oscillator that accounts for the main anomalies like the
\textquotedblleft waterfall\textquotedblright\ and the large width of the
ARPES tails, is the damping parameter. We further find quantitative
similarities between the parameters of the oscillator and data available
from optical experiments on different high Tc materials, namely, the
incoherent excitation spectrum in the MIR region. It is known that this
mid-gap incoherent part of the spectral weight emerges more quickly than the
"coherent" Drude part upon doping \cite{Nakano}. Unfortunately, optical data
can be related only to the long wave limit ($q=0$) of the \textquotedblleft
overdamped boson\textquotedblright , as there are no experimental probes
available to study its dispersion characteristics. Therefore for our model
calculations we have neglected the dispersion of this MIR oscillator.
Nevertheless, we discuss some features reported in experiments that can be
accounted for by considering an anisotropic dispersion.

In the Table I we have summarized the experimental data for the Lorentz
oscillator component in the relevant MIR region of energies extracted from
optical conductivity for the Drude-Lorentz representation of the dielectric
function
\begin{equation}
\epsilon (\omega )=\epsilon _{\infty }+\sum_{k=1}^{N}\frac{\omega _{p,k}^{2}%
}{\omega _{0,k}^{2}-\omega ^{2}-i\omega \gamma _{k}}.  \label{DF}
\end{equation}%
The Drude component with damping $\gamma _{1}$ and weight determined by the
plasmon frequency $\omega _{p,1}$ ($\sim 1.5-2$ eV) is centered at $\omega
=0,$ the frequency and damping of the discussed excitation are further
denoted by $\omega _{0}$ and $\gamma ,$ the \textquotedblleft
effective\textquotedblright\ frequency $\Omega $ is explained below. The
values of the oscillator strength corresponding to $\omega _{0}$ can be
found in the listed papers, but are generally comparable to the weight of
the Drude part of the dielectric function at optimal doping. It should be
noted that data in the table are both doping and temperature dependent and
that respective multi-component analysis in terms Drude-Lorentz oscillators
depends on the number of components chosen. In particular, physical
arguments suggest that an oscillator in the phononic frequency range of the
MIR should be distinguished from the oscillator singled out in the table.
Such a fitting procedure can be formalized, for instance, in the
Kramers-Kronig constrained variational analysis, e.g. \cite{Heumen, Kuzmenko}%
. The obvious deficiency of such analysis is the absence of frequency
dependence in the mass and relaxation rate of the Drude component
(generalized Drude model) \cite{DordevicBasov}. However for the energies of
interest in the present context it is not expected to dramatically alter the
estimations. Indeed, the generalized Drude model is appropriate for the
lower energy part of the spectrum characterized by a single scattering
mechanism, but is insufficient to account for the higher energy scale
related to the HEA. For the estimates discussed in the present paper we
assume that these scales can be treated independently in calculating the
electronic selfenergy. As already mentioned, the long wavelength values of
characteristic frequencies in the Table should be viewed as an estimate for
the actual frequency of the overdamped charge excitation for our model since
its value at finite momentum is not known.

We introduce the retarded bosonic propagator $B\left( q,\omega \right) $ in
the form of damped oscillator \cite{Mahan} with spectral density $\rho
_{B}\left( q,\omega \right) $:
\begin{equation}
B\left( \omega \right) =\frac{a_{q}\omega _{q}}{\omega ^{2}-\omega
_{q}^{2}+i\gamma _{q}\omega },\rho _{B}\left( \omega \right) =\frac{\omega
\gamma _{q}a_{q}\omega _{q}/\pi }{\left( \omega ^{2}-\omega _{q}^{2}\right)
^{2}+\left( \gamma _{q}\omega \right) ^{2}}.  \label{BSD}
\end{equation}%
The choice of (\ref{BSD}) in our model is motivated by its compliance with
the generic form of the response functions frequently used in analyzing
experimental data that, in turn, corresponds to physical requirements, such
as Kramers-Kronig relations. Coupling to electron excitations is considered
in the second order of perturbation theory, i.e. non-selfconsistently. The
scale of the boson frequency is assumed to still allow the adiabatic
approximation to be reasonable at least for evaluation purposes (see \cite%
{Danylenko} for a recent discussion). Although the vertex corrections and
selfconsistent treatment are beyond the scope of the paper, their analysis
certainly deserves a special consideration. Using the spectral
representation for the Matsubara Green function, after carrying out the
frequency summation and analytic continuation to real frequencies one
obtains the second order expression for the retarded electron selfenergy,
e.g. \cite{Marsiglio, Dogan},%
\begin{equation}
\Sigma _{R}\left( k,\omega \right) =\frac{1}{N}\sum_{q}g^{2}\left(
k,q\right) \int_{-\infty }^{\infty }\int_{-\infty }^{\infty }\rho _{F}\left(
k-q,x\right) \rho _{B}\left( q,y\right) \frac{n_{B}\left( y\right)
+n_{F}\left( x\right) }{\omega +y-x+i0^{+}}dxdy,  \label{sigma}
\end{equation}%
where $g\left( q\right) $ is the coupling constant and $n_{B,F}$ are Boson,
Fermion distribution functions. The linewidth of the electronic spectral
function $\rho _{F}\left( k-q,x\right) $ is neglected, $\rho _{F}\left(
k-q,x\right) =\delta \left( x-\varepsilon _{k-q}\right) ,$ and a tight
binding parametrization of the band dispersion is used as explained above,
that allows to focus on the scattering mechanism due to the
\textquotedblleft boson\textquotedblright . If also the linewidth of $\rho
_{B}\left( \omega \right) $ is neglected, the expression (\ref{sigma})
transforms into the one that has been often employed in describing the LEK (
see, e.g.\cite{ZhouCuk}) and results in the usual logarithmic singularity of
Re$\Sigma \left( \omega \right) $ at the boson frequency. To study the
effect of the broad linewidth on electron spectrum we neglect the dispersion
in spectral function (\ref{BSD}) and in the coupling $g\left( q\right) .$
Then the selfenergy becomes a local function%
\begin{equation}
\Sigma _{R}\left( \omega \right) =\lambda _{c}\gamma \omega
_{0}^{2}\int_{-\infty }^{\infty }dy\frac{y}{\left( y^{2}-\omega
_{0}^{2}\right) ^{2}+\left( \gamma y\right) ^{2}}\int dE\rho _{0}\left(
E\right) \left( \frac{n_{B}\left( y\right) +n_{F}\left( E\right) }{\omega
+y-E+i0^{+}}\right) ,  \label{sigmac}
\end{equation}%
where $\rho _{0}\left( E\right) $ is the electronic density of states, $%
\lambda _{c}$ is the dimensionless coupling constant. The coupling constant $%
\lambda $ determined from the slope of Re $\Sigma \left( \omega \right) $ is
slightly different from $\lambda _{c}$. The band width of $\varepsilon _{k}$
is taken as the energy unit in numerical calculations.

\begin{figure}[tbph]
\centering
\includegraphics[height=5cm,width=6cm]{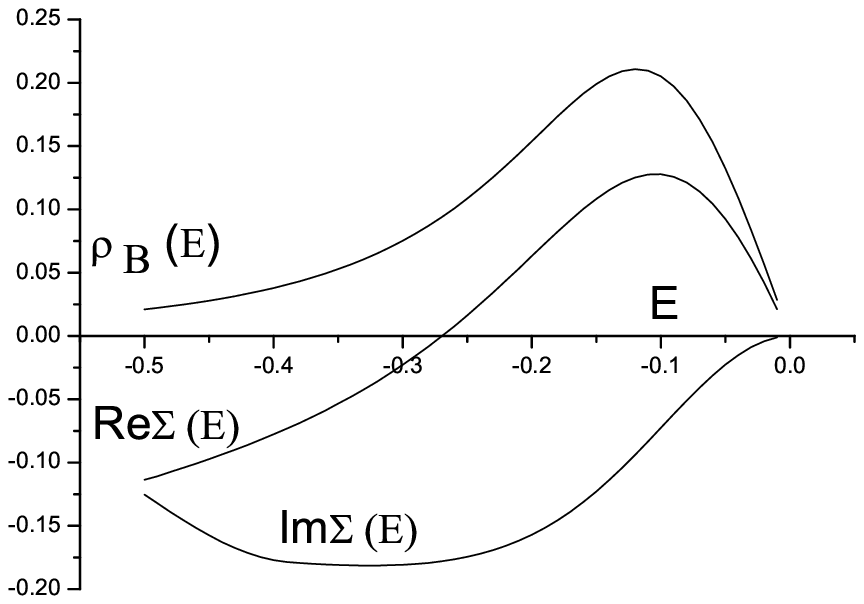}
\caption{Real and imaginary parts of the selfenergy $\Sigma \left( \protect%
\omega \right) $ (\protect\ref{sigmac}) resulting from interaction with the
overdamped oscillator with spectral density $\protect\rho _{B}\left( \protect%
\omega \right) $. Parameters are $\protect\omega _{0}=0.18,\protect\gamma %
=0.3,\protect\lambda=1.3,T=0.001,\protect\delta =0.15.$ The flattening of Im$%
\Sigma \left( \protect\omega \right) $ corresponds to the incoherent
"waterfall" region in ARPES with almost constant linewidth. }
\label{fig2}
\end{figure}

The spectral density of the damped oscillator (\ref{BSD}) is peaked at the
frequency $\Omega =\left( 12-6R^{2}+6\sqrt{16-4R^{2}+R^{4}}\right)
^{1/2}\times \omega _{0}/6,$ where $R=\gamma /\omega _{0}$. It is clear that
the effective frequency $\Omega $ is significantly lower than $\omega _{0}$
due to a large damping typical for the MIR oscillator. As seen from the Fig.%
\ref{fig2}, it is this effective frequency that sets the location of the
high energy kink. In the figure we plot the Re$\Sigma \left( \omega \right) $
and Im$\Sigma \left( \omega \right) $ together with the \textquotedblleft
boson\textquotedblright\ spectral function $\rho _{B}\left( \omega \right) $
to make the correlation between the peaks clearly visible. The slope of the
Re$\Sigma \left( \omega \right) $ corresponds to the coupling constant $1.3$
and the ratio of $t^{\prime }/t=0.1$ was taken to model the LSCO band $%
\varepsilon _{k}$. One can also see that the $\left( -Im\Sigma \left( \omega
\right) \right) $ follows a $\omega ^{2}$ at low energies and its inflection
point corresponds to $\Omega .$ After reaching its maximum at higher energy $%
\left( -Im\Sigma \left( \omega \right) \right) $ follows an almost flat
region until it starts decreasing closer to the band bottom. Such behavior
has been reported in \cite{Graf, Chang} for the linewidth of the ARPES
spectral function. In our calculations we clearly see the feature mentioned
in \cite{HwangNicol} and related to the finite bandwidth when the
renormalized dispersion curve crosses with bare one. The crossing occurs at
certain nonzero frequency within the waterfall region where Re$\Sigma \left(
\omega \right) =0.$

\begin{figure}[tbph]
\centering
\includegraphics[height=5cm,width=5cm]{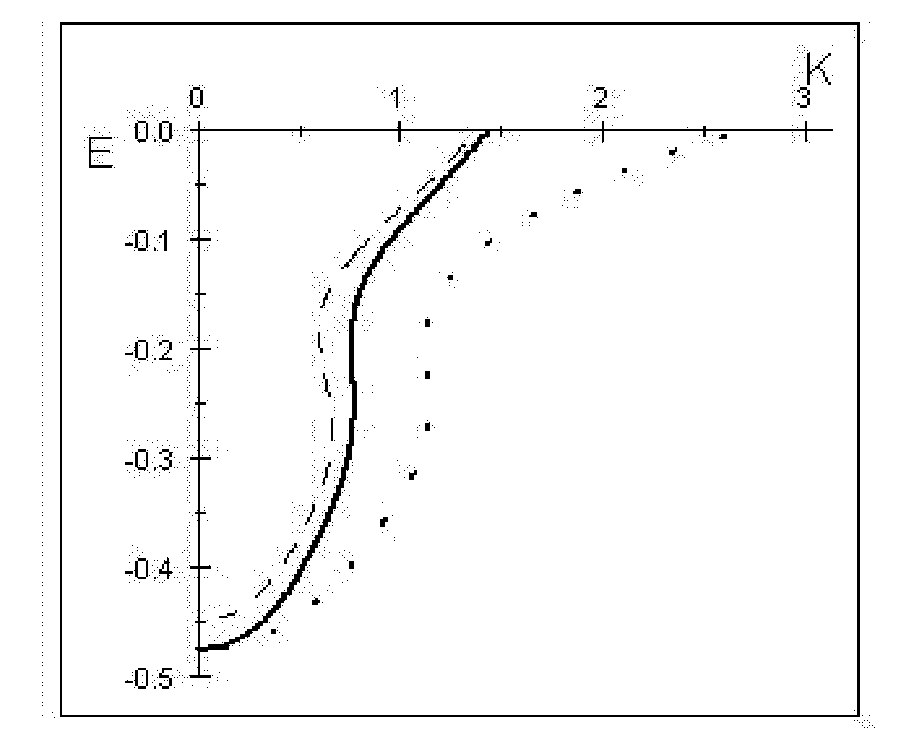}
\caption{Dispersion of the spectral density maximum along the diagonal
direction, $K=Kx=Ky$, (continuous line $\protect\delta =0.15,$ dashed line $%
\protect\delta =0.25$) and orthogonal direction, $K=Kx(Ky),Ky(Kx)=0$,
(dotted line $\protect\delta =0.15$) of the Brillouin zone. Parameters are: $%
\protect\omega _{0}=0.18,\protect\gamma =0.21,\protect\lambda =1.3,T=0.001.$
}
\label{fig3}
\end{figure}

\begin{figure}[tbph]
\centering
\includegraphics[height=5cm,width=5cm]{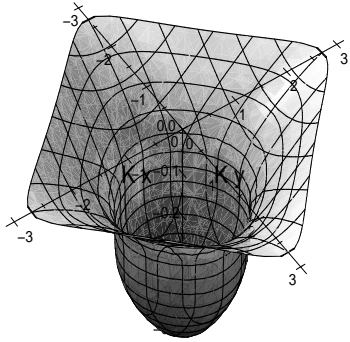}
\caption{Three dimensional plot of the surface defining the position of the
peak in the spectral density as a function of energy and in-plane momentum.
Parameters correspond to Fig.\protect\ref{fig3}.}
\label{fig4}
\end{figure}

The calculated dispersion of the spectral function peak along the two main
symmetry directions of the Brillouin zone are presented in Fig.\ref{fig3}
for doping $\delta =0.15$ and band structure of BISCO 2212. The dashed curve
shows the evolution of the curve at larger doping, $\delta =0.25,$ for the
diagonal direction. The HEA is close to $k_{0}\simeq \pi /4$ at $E_{1}\simeq
\Omega =0.15$ ($\simeq 0.4$ eV), the \textquotedblleft
waterfall\textquotedblright\ is superceded by parabolic dispersion at $%
E_{2}\simeq 0.33$ ($\simeq 0.85\,$eV$\allowbreak $). The momentum of HEA $%
k_{0}$ decreases with doping. These results qualitatively agree with
experimental data \cite{Graf, Valla} and appear to suggest that the band
structure is of primary importance for the location and doping dependence of
the HEA, as the doping dependence of $k_{0}$ follows the shrinking of the
Fermi surface. However, it should be taken into account that also the
parameters of the MIR oscillator depend on doping. In particular, the
quantitatively weaker dependence of $k_{0}$ observed in experiments could be
related to the decrease of $\omega _{0}$ with doping. We are not aware of
the doping dependence of the linewidth of the spectral function in the
waterfall region, but our calculation (not shown) indicates its decrease
with doping. The temperature dependence of $\Sigma \left( \omega \right) $
is very weak in the relevant region of energies. We find a broader linewidth
in the $\left( 0,1\right) $ and $\left( 1,0\right) $ directions as compared
to the diagonal in agreement with the measurements presented in \cite{Pan}.
Fig.\ref{fig4} shows the calculated three dimensional dispersion of the
spectral density peak within the Brillouin zone obtained by neglecting the
dispersion of the damped oscillator that agrees with the results reported in
\cite{Graf}. However, as noted, the ARPES measurements presented in some
other papers \cite{KordyukBorisenko, Valla, Chang} show an anisotropic
structure of the HEA. These results can be described within the present
model by introducing the anisotropy into the parameters of the oscillator.
For instance, the difference in energy of the kink for momentum cuts along
different directions of the Brilouin zone found in \cite{Chang} can be
explained by a larger value of $\Omega $ along the diagonal direction. In a
different context, it has been noted that the MIR peak in optics is affected
by lattice oscillations \cite{Piekarz}. It has also been suggested \cite{CCM}
that the anomalous softening of the bond-stretching in plane phonon mode is
related to an overdamped charge excitation in the same MIR part of the
spectrum and that the behavior of this mode could serve as an
energy-momentum resolved probe for the charge excitation. This raises an
intriguing possibility that the same overdamped excitation is responsible
for the phonon anomaly and the HEA. One of its characteristics implied from
the phonon analysis is the anisotropy of the charge response at finite
momentum in the energy range that matches the grid-like pattern mentioned in
\cite{KordyukBorisenko}. Namely that anomalous phonon softening occurs in
the $\left( 0,1\right) ,\left( 1,0\right) $ directions and a much smaller
effect is observed in the $\left( 1,1\right) $ direction. This anisotropy is
related to a larger charge response located at lower energies for the
orthogonal direction as compared to more dispersive and weaker response
along the diagonal. The strength of the anisotropy depends on the material
and doping. Interestingly, in the papers \cite{Valla, Chang} the energy of
the kink (in absolute value) in $LBCO,$ $BISCO$ $2212$ was found to decrease
significantly away from the diagonal direction of the Brillouin zone. This
behavior is also in agreement with the MIR charge \textquotedblleft
boson\textquotedblright\ scenario proposed here. The broad incoherent
spectrum above the high energy kink emerges due to the composite nature of
electronic excitation (in itself a source of finite linewidth) with emission
of an \textquotedblleft overdamped boson\textquotedblright . This explains
the qualitative difference between the high and low energy kinks. In the
latter case the linewidth of the boson is not more than $10\%$ of its energy
scale, while in the case of MIR it is an order of magnitude larger$.$ At the
same time, for a comparable coupling constant, such linewidth accounts for
the verticality of the dispersion and linewidth of the waterfall. As we have
seen, the Im$\Sigma $ remains almost flat in this region. At energies around
$0.9$ eV the Im$\Sigma $ starts to decrease and the dispersion recovers the
bare band parabolic shape, albeit shifted to higher energies. It should be
mentioned that the optical conductivity data have served only for
estimations and, as discussed above, one needs to have a probe for momentum
resolved charge excitations in the $0.4-0.5$ eV range to characterize the
\textquotedblleft overdamped boson\textquotedblright .

In [\onlinecite{Graf}] several other types of materials have been
scrutinised for HEA, but no evidence has been found. It follows from the
proposed scenario that HEA should exist in the ARPES of other perovskite
materials where the incoherent MIR excitation has been observed (and vice
versa). For instance, in the paper \cite{Sun} the ARPES analysis is focused
on the LEK at \ $60$ meV in bilayer manganite $La_{2-2x}Sr_{1+2x}Mn_{2}O_{7}.
$ However, also the data for higher energies are presented, where the
existence of the other kink at around $0.45$ eV is clearly visible. It would
also be interesting to carry out a similar search for the HEA in ARPES
experiments on a non-magnetic perovskite such as $Ba_{1-x}K_{x}BiO_{3}$ with
a relatively high Tc that is also known to have an incoherent charge
excitation in the MIR about $0.4-0.6$ eV \cite{Kaufmann, Ahmad, Puchkov} as
well as anomalous softening of the bond-stretching phonon mode \cite{Braden}.


\begin{table}[tbp]
\caption{Parameters of the MIR oscillator in (\protect\ref{DF}) determined
from the Drude-Lorentz analysis of the optical conductivity.}
\begin{center}
\begin{tabular}{|c|c|c|c|}
\hline
material & $\omega _{0}$ & $\gamma $ & $\Omega $ \\ \hline
Quijada,\cite{Quijada}\ $Bi_{2}Sr_{2}CaCu_{2}O_{8},$a axis & $0.545$ & $0.95$
& $0.358$ \\ \hline
Quijada,\cite{Quijada}\ $Bi_{2}Sr_{2}CaCu_{2}O_{8},$b axis & $0.51$ & $1$ & $%
0.3$ \\ \hline
Wang, \cite{WangZheng}$\ La_{1.9}Ca_{1.1}Cu_{2}0_{6+\delta }$ & $0.59$ & $%
0.74$ & $0.47$ \\ \hline
Wang,\cite{WangZheng}$\ La_{1.85}Sr_{0.15}CaCu_{2}0_{6+\delta }$ & $0.57$ & $%
0.67$ & $0.47$ \\ \hline
Uchida, \cite{Uchida} $La_{1.9}Sr_{0.1}CuO_{4+\delta }$ & $0.46$ & $0.8$ & $%
0.3$ \\ \hline
Kircher,\cite{Kircher} $YBa_{2}Cu_{4}O_{8}$ & $0.55$ & $1.12$ & $0.31$ \\
\hline
Heumen,\cite{Heumen} $HgBa_{2}CuO_{4+\delta }$ & $0.48$ & $0.9$ & $0.3$ \\
\hline
Carbone, \cite{Carbone}\ $Bi_{2}Sr_{2}Ca_{2}Cu_{3}O_{10}$ & $0.63$ & $1$ & $%
0.44$ \\ \hline
Orenstein, \cite{Orenstein} $YBa_{2}Cu_{3}O_{6.9}$ & $0.65$ & $1.3$ & $0.38$
\\ \hline
\end{tabular}%
\end{center}
\end{table}

\end{document}